\documentstyle[mprocl]{article}
\begin{document}

\title{FROM TAUB NUMBERS TO THE BONDI MASS\footnote{\
To appear in Class. Quan. Grav. {\bf 14}, 1899 (1997)}}

\author{E.N. Glass\footnote{Permanent\
 address: Physics Department, University of Windsor, Ontario, CANADA}}
\address{Physics Department, University of Michigan, Ann Arbor, MI, USA}

\maketitle

\begin{abstract}
Taub numbers are studied on asymptotically flat backgrounds with Killing
symmetries. When the field equations are solved for a background spacetime
and higher order functional derivatives (higher order variational
derivatives of the Hilbert Lagrangean) are solved for perturbations from the
background, such perturbed space-times admit zeroth, first, and second order
Taub numbers. Zeroth order Taub numbers are Komar constants (upto numerical
factors) or Penrose-Goldberg constants of the background. For a Killing
symmetry of the background, first order Taub numbers give the contribution
of the linearized perturbation to the associated backgound quantity, such as
the perturbing mass. Second order Taub numbers give the contribution of
second order perturbations to the background quantity. The Bondi mass is a
sum of first and second order Taubs numbers on a Minkowski background.
\end{abstract}

\section{INTRODUCTION}

To define the Bondi mass one needs an asymptotically flat manifold and the
notion of future null infinity $\cal{I}^{+}$. The Bondi metric\cite
{bondi} is axisymmetric and has 4 metric functions $V,$ $U,$ $\beta ,\gamma
. $
\\ \\
$g_{\mu \nu }^{Bondi}dx^\mu dx^\nu =(Ve^{2\beta }/r-r^2U^2e^{2\gamma
})du^2+2e^{2\beta }dudr+2r^2Ue^{2\gamma }dud\theta $

$\ \ \ \ \ \ \ \ \ \ \ \ \ \ \ \ -r^2(e^{2\gamma }d\theta ^2+e^{-2\gamma
}\sin ^2\theta d\phi ^2)$
\\ \\
where $\gamma =O(1/r),\ \beta =O(1/r^2),\ U=O(1/r^2)$. The asymptotic form
of metric function $V$ is $V=r-2M(u,\theta )+O(1/r).\ $\ $M(u,\theta )$ is
the Bondi mass aspect 
\[
-2M(u,\theta )=\Psi _2^0+\bar{\Psi}_2^0+\partial _u(\sigma ^0\bar{\sigma}^0).
\]
The Bondi mass is the 2-surface integral of the mass aspect over a
topological 2-sphere at $\cal{I}^{+}$%
\[
M_{Bondi}=-\frac 1{8\pi }\int\limits_{S^2}[\Psi _2^0+\bar{\Psi}%
_2^0+\partial _u(\sigma ^0\bar{\sigma}^0)]\,d\Omega 
\]

\section{EARLY CALCULATIONS OF THE BONDI MASS}

The first computation of the Bondi mass was done by J.N. Goldberg\cite{josh}
with use of the Einstein pseudotensor and von Freud superpotential: $(\sqrt{%
-g}Einstein)_\alpha ^{\ \beta }=\partial _\nu (superpotential)_\alpha ^{\
[\beta \nu ]}+(pseudotensor)_\alpha ^{\ \beta }.$

Calculating in an asymptotically Cartesian frame and transforming to null
spherical coordinates, Goldberg integrated the superpotential over an
asymptotic 2-sphere at future null infinity and, using the generator of
asymptotic time translations, obtained the Bondi mass.

Winicour and Tamburino \cite{jeffandlou} constructed a tensorial calculation
by modifying the Komar superpotential. For the null surface $u=const.$ they
added a term which eliminated off-surface derivatives. Using an asymptotic
symmetry, the integral of the modified Komar superpotential, called a
''linkage'', yields the Bondi mass at $\cal{I}^{+}$. Unfortunately the
linkage construction does not arise from a variational principle.

\section{TAUB METHOD}

The Taub method is fully developed in \cite{taub1,taub2,einmax}. We write
the Bondi metric in a perturbation expansion as $g_{\alpha \beta
}^{Bondi}(\lambda )=\eta _{\alpha \beta }+\lambda h_{\alpha \beta
}^{(1)}+\lambda ^2h_{\alpha \beta }^{(2)}+\cdots $. To insure a single
coordinate system for each tensor in the perturbation expansion, the
Minkowski background metric is constructed from the Bondi metric. We solve
the Newman-Penrose form of Einstein's equations with initial data $\Psi
_0(u_0,r,\theta )=\Psi _1^0(u_0,\theta )=\Psi _2^0(u_0,\theta )=\sigma
^0(u,\theta )=0.$ One obtains Bondi metric functions $V=r,\ U=\beta =\gamma
=0$, i.e. the Minkowski metric. The symmetry generator of mass, the
backgound timelike Killing vector, $k_t=\partial _u,$\ is covariant
constant.  The Taub superpotential $U_{Taub}^{\alpha \beta
}(k_t,h)=(-g)^{\frac 12}k_t^\nu (\delta _\nu ^{\ [\alpha }h_{\ \ \mu
}^{\beta ]\ ;\mu }-\delta _\nu ^{\ [\alpha }h^{;\beta ]}+h_\nu ^{\ [\alpha
;\beta ]})$ is used to compute Taub numbers where the n$^{th}$ Taub number
is $\tau _n(k_t,h^{(n)})=-\frac 1{8\pi }\int\limits_{\partial \cal{N}}
U_{Taub}^{\alpha \beta }(k_t,h^{(n)})\ dS_{\alpha \beta }$. Each of the $
h^{(n)}$ and hence each of the $\tau _n$ derives from a variational
derivative of\ the Hilbert action, \ the $n^{th}$ number from the $n+1^{th}$
variation.

The Bondi mass results from the sum of first and second numbers: $%
M_{Bondi}=\tau _1(k_t,h^{(1)})+\tau _2(k_t,h^{(2)})$. $\tau _1$ has the
non-radiative mass contribution and $\tau _2$ contains the news function.

\section{PERTURBATION CALCULATION}

To find $\tau _1\ $solve \ $DEin(\,h^{(1)}\,)=0$ \ for\ $h_{\mu \nu }^{(1)}$%
\ using the Newman-Penrose equations on the Minkowski background with tetrad
\ $l=du,\ n=\frac 12du+dr,\ m=-(r/\sqrt{2})(d\theta +isin\theta d\phi )$.
Find the spin coefficients, metric components, and Weyl tensor components
with initial data \ \{$\Psi _0(u_0,r,\theta ),\ \Psi _1^0(u_0,\theta ),\
\Psi _2^0(u_0,\theta ),\ \sigma ^0(u,\theta )$\}.
\\ \\
$h_{\mu \nu }^{(1)}=[(\Psi _2^0+\bar{\Psi}_2^0)/r]l_\mu l_\nu -(2\bar{\sigma}%
^0/r)m_\mu m_\nu -(2\sigma ^0/r)\bar{m}_\mu \bar{m}_\nu $

$\ \ \ -(\partial \mbox{$\hspace{-2mm} /$}\bar{\sigma}^0/r)(l_\mu m_\nu +m_\mu
l_\nu )-(\bar{\partial}\mbox{$\hspace{-2mm} /$}\sigma ^0/r)(l_\mu \bar{m}%
_\nu +\bar{m}_\mu l_\nu ).$
\\ \\
Substitute into \ $U_{Taub}^{\alpha \beta }(k_t,h^{(1)})\ $and obtain $\tau
_1=-\frac 1{8\pi }\int\limits_{S^2}(\Psi _2^0+\bar{\Psi}_2^0)\ d\Omega .$
This is $M_{Bondi}$ when the news $\partial _u\sigma ^0=0.$ Vacuum
Schwarzschild has $\sigma =0,\ \Psi _2^0=-m\ $and so$\ M_{Bondi}=\tau _1=m.$

To find $\tau _2\ $iterate the Newman-Penrose equations. The second order
Bianchi identities for the Weyl tensor components have first order
quantities as sources whereas the first order equations were source-free 
\cite{ctjn}.
\\ \\
$U_{Taub}^{\alpha \beta }(k_t,h^{(2)})=\sqrt{-g}\,l^{[\alpha }n^{\beta ]}[-2%
\frac{\partial _u(\sigma ^0\bar{\sigma}^0)}{r^2}+O_3].$
\\ \\
$\tau _2(k_t,h^{(2)})=-\frac 1{8\pi }\int\limits_{S^2}[\partial _u(\sigma
^0\bar{\sigma}^0)]\,d\Omega .$
\\ \\
$M_{Bondi}=\tau _1+\tau _2.$\ \ Higher orders fall off faster and do not
contribute to the Bondi mass

\end{document}